\documentclass[prb,aps,reprint,superscriptaddress,amsmath,amssymb,longbibliography]{revtex4-2}

\usepackage{graphicx,tabularx,makecell}
\usepackage{amsmath,amsfonts,amssymb}
\usepackage{mathtools}
\usepackage{bm,dsfont}
\usepackage[dvipsnames]{xcolor}
\usepackage[bookmarks=true,colorlinks=true,urlcolor=blue,linkcolor=blue,citecolor=blue,breaklinks]{hyperref}
\usepackage[all]{hypcap}
\usepackage[T1]{fontenc}
\usepackage[utf8]{inputenc}
\usepackage{xspace}
\usepackage{ulem}
\usepackage{listings}
\usepackage{xcolor}

\definecolor{codegreen}{rgb}{0,0.6,0}
\definecolor{codegray}{rgb}{0.5,0.5,0.5}
\definecolor{codepurple}{rgb}{0.58,0,0.82}
\definecolor{backcolour}{rgb}{0.95,0.95,0.92}

\lstdefinestyle{mystyle}{
    backgroundcolor=\color{backcolour},   
    commentstyle=\color{codegreen},
    morekeywords={dot,Cross2,G,phase,vtotal,sp}
    keywordstyle=\color{magenta},
    numberstyle=\tiny\color{codegray},
    stringstyle=\color{codepurple},
    basicstyle=\ttfamily\footnotesize,
    breakatwhitespace=false,         
    breaklines=true,                 
    captionpos=b,                    
    keepspaces=true,                 
    numbers=left,                    
    numbersep=5pt,                  
    showspaces=false,                
    showstringspaces=false,
    showtabs=false,                  
    tabsize=2
}
\graphicspath{{figures/}}

\newcommand{\mytitle}{Lattice Relaxation in Moir\'e Heterobilayers}

\begin{document}

\title{\mytitle}

\author{Christophe De Beule}
\affiliation{Department of Physics, University of Antwerp, Groenenborgerlaan 171, 2020 Antwerp, Belgium}
\author{Yiyang Lai}
\affiliation{Department of Physics, Washington University in St. Louis, St. Louis, Missouri 63130, United States}
\author{Liangtao Peng}
\affiliation{Department of Physics, Washington University in St. Louis, St. Louis, Missouri 63130, United States}
\author{Daniel Bennett}
\affiliation{School of Electrical and Electronic Engineering, Nanyang Technological University Singapore, 50 Nanyang Avenue, 639798, Singapore}
\author{Shaffique Adam}
\affiliation{Department of Physics, Washington University in St. Louis, St. Louis, Missouri 63130, United States}
\affiliation{Institute of Materials Science and Engineering, Washington University in St. Louis, St. Louis, Missouri 63130, USA}

\date{\today}

\begin{abstract}
We develop an analytical theory for lattice relaxation in twisted moir\'e heterobilayers, accounting for lattice mismatch, twist, external biaxial heterostrain, and different elastic constants. Starting from continuum elasticity, we derive the self-consistent equations for the in-plane displacement fields and obtain simple perturbative expressions for the layer-resolved in-plane displacement fields induced by lattice relaxation. We apply our theory to graphene on hBN and representative 2H transition metal dichalcogenide heterobilayers, including MoTe$_2$/WSe$_2$ and WSe$_2$/WS$_2$. Our analytical results agree very well with full numerical solutions over experimentally relevant parameters. We further show that heterobilayers can exhibit a buckling instability near alignment, driven by compressive in-plane strain due to moir\'e relaxation. Our results provide a simple theoretical framework for incorporating lattice relaxation in realistic moir\'e heterostructures.
\end{abstract}

\maketitle

\section{Introduction}

Van der Waals heterostructures provide a versatile platform for discovering both new physics~\cite{balentsNP2020,andreiNM2020,makNN2022} and novel technologies \cite{novoselovS2016,andreiNRM2021,kennesNP2021}. Already more than 10 years ago it was realized both experimentally \cite{woodsNP2014} and theoretically \cite{jungNC2015,jungPRB2017} that stacked atomically thin two-dimensional (2D) materials behave more like elastic membranes than rigid crystals. Hence, the structural and electronic properties of moir\'e superlattices, formed by stacking layers with a relative twist or lattice mismatch, can be significantly modified by lattice relaxation. In particular, this gives rise to important new physics, including the separation of the flat bands from dispersive bands in magic-angle twisted bilayer graphene \cite{namPRB2017,caoN2018} and the opening of a topological gap in graphene on hexagonal boron nitride (hBN) \cite{jungNC2015,ametPRL2013,huntS2013}. This stabilizes an anomalous Hall phase \cite{bultinckPRL2020} resulting in the observation of the orbital anomalous Hall effect \cite{sharpeS2019,serlinS2020}.

Atomic reconstruction in moir\'es is controlled by the competition between intralayer elastic energy and interlayer stacking energy \cite{woodsNP2014,jungNC2015,carrPRB2018}, becoming increasingly important as the moir\'e period increases. Much of the work on lattice relaxation over the past decade has been heavily numerical, either by solving the fully relaxed density functional theory \cite{lecontePRB2022,naikPRL2018, cantelePRR2020,lucignanoPRB2019} or molecular dynamics simulations \cite{lecontePRB2022}. These methods become unfeasible for small twist angles due to the large number of atoms in the moir\'e unit cell ($\sim 1 / \theta^2$) and provide limited physical insights into the relaxation process. However, very recently there have emerged analytical treatments of moir\'e reconstruction, including a large-angle theory for both monoatomic hexagonal lattices \cite{ezziPRL2024,kangPRB2025} (e.g.\ twisted bilayer graphene) and binary hexagonal lattices [e.g.\ transition metal dichalcogenides (TMDs)] \cite{ezziPRL2024,yu2025} that agree well with available numerics. Even more recently, we established an approximate analytical treatment \cite{debeulea2025} of lattice relaxation for marginal twist angles, matching predictions from ab initio methods \cite{zhangNC2024} and numerical solutions of continuum elasticity \cite{namPRB2017,carrPRB2018,bennettPRB2022}. In particular, we find an emergent universality where the theory is characterized by a single twist-angle dependent parameter that captures both the small and large angle regimes. It enables accurate predictions of the acoustic displacement fields of any homo moir\'e interface built from layers with $D_{3h}$ symmetry (twist near $0^\circ$) or $D_{3d}$ (twist near $60^\circ$) and encompasses a vast number of moir\'e materials of contemporary interest. Examples of $D_{3h}$ homobilayers are: twisted bilayer graphene, 2H TMDs (MoS$_2$, WSe$_2$, MoTe$_2$, WS$_2$, NbSe$_2$, etc.), and 2D magnets like 2H Fe$_3$GeTe$_2$. Some examples of $D_{3d}$ homobilayers are given by: twisted double bilayer graphene, 1T TMDs (ZrS$_2$, HfS$_2$, HfSe$_2$, TiS$_2$, etc.), and 2D magnets e.g.\ NiI$_2$, MnI$_2$, and FeCl$_2$, and CrX$_3$ systems such as CrI$_3$.

However, the previous work focused only on homobilayers, where the two twisted sheets are made from the same material. Heterobilayers, comprising twisted sheets of two different materials, are equally interesting: for example, it is well-known that graphene on hBN features a Hofstadter’s butterfly and the fractional quantum Hall effect \cite{deanN2013,ponomarenkoN2013} and strong moir\'e potentials \cite{kleinN2026}. Also, in Ref.\ \cite{liN2021} a topological phase transition from a Mott insulator to a quantum anomalous Hall insulator was observed in MoTe$_2$/WSe$_2$ moir\'e heterobilayers and Ref.\ \cite{reganN2020} observed Mott and generalized Wigner crystal states in WSe$_2$/WS$_2$ moir\'e superlattices. And theoretically, the formation of quantum dots localized at the highly strained nodes of domain wall networks in MoX$_2$/WX$_2$ heterostructures have been proposed \cite{solteroNL2024}.

The goal of this paper is to generalize the analytical framework \cite{ezziPRL2024,debeulea2025} to heterobilayers where lattice relaxation is expected to be important even without twisting when the lattice mismatch between inequivalent layers is small. Here, we successfully develop a one-shot perturbation theory for the in-plane acoustic displacement fields that works for any heterobilayers with the same Bravais lattice allowing for lattice mismatch between layers with different in-plane point group symmetry, e.g.\ for graphene on hBN: graphene has $C_{6v}$ symmetry, while hBN has $C_{3v}$ symmetry, with approximately $2 \%$ mismatch between lattice constants. We apply our analytical theory to several moir\'e systems of interest such as graphene on hBN and 2H TMDs, including MoTe$_2$/WSe$_2$, WS$_2$/WSe$_2$, and WS$_2$/MoSe$_2$, and compare our work with numerical results. In general, the analytical theory is accurate within $10 \%$ for any twist angle. Beyond the in-plane results, we also discuss the role of out-of-plane displacements in free-standing heterobilayers, and  identify a buckling transition driven by moir\'e reconstruction. Our current work extends the analytical framework of lattice relaxation to a much larger class of Van der Waals heterostructures.  

\begin{figure*}
    \centering
    \includegraphics[width=\linewidth]{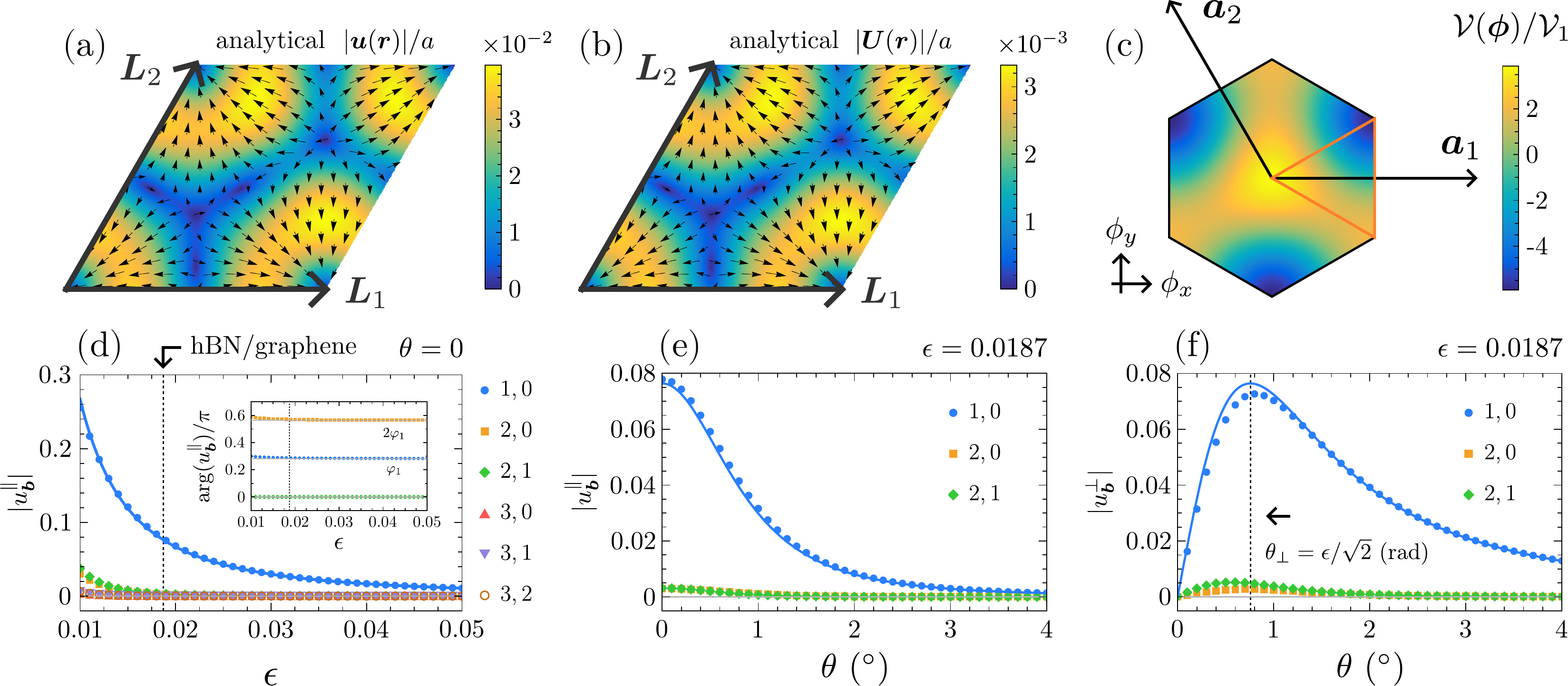}
    \caption{First-star analytical theory for (a) the relative in-plane displacement field $\bm u = \bm u_1 - \bm u_2$ and (b) the layer center-of-mass field $\bm U = \bm u_1 + \bm u_2$, for graphene on hBN due to lattice relaxation. Here, the magnitude is shown in units of the average lattice constant $a = (a_1 + a_2)/2$ with $a_1 > a_2$. (c) First-star stacking potential with $C_{3v} = \left< \mathcal C_{3z}, \mathcal M_x \right>$ symmetry in configuration space from Eq.\ \eqref{eq:stacking} for $\varphi_1 = 50^\circ$. (d) Longitudinal Fourier components of $\bm u$, where the dots are numerical solutions of the self-consistent equations given in Eqs.\ \ref{eq:sc1} and \ref{eq:sc2}, and the curves give the one-shot result. The legend indicates the star following the notation of Table \ref{tab:tab1}. Shown as a function of lattice mismatch $\epsilon$ for $\theta = 0$. (e) Longitudinal and (f) transverse components versus twist angle for $\epsilon = 0.0187$. Components that are not shown are negligible. Fourier components of the layer center-of-mass field are obtained from $U_{\bm b}^\parallel = A^\parallel u_{\bm b}^\parallel$ and $U_{\bm b}^\perp = A^\perp u_{\bm b}^\perp$ with $A^\parallel \approx A^\perp \approx 0.08$. Parameters used are shown in Tables \ref{tab:tab2} and \ref{tab:tab3}.}
    \label{fig:fig1}
\end{figure*}

\section{Structural relaxation in moir\'e bilayers}

We start from the total structural energy of the bilayer system, which is composed of elastic energy, which favors the rigid moir\'e without deformations, and adhesion energy, which favors deformations that increase regions with energetically-favorable layer stacking. Treating the layers as isotropic continuous membranes, the elastic energy is given by \cite{landau1986}
\begin{equation} \label{eq:elastic}
    H_\text{elas} = \frac{1}{2} \sum_{l=1,2} \int d^2 \bm r \left[ \lambda_l u_{ii}^{(l)} u_{ii}^{(l)} + 2 \mu_l u_{ij}^{(l)} u_{ji}^{(l)} \right],
\end{equation}
where the sum runs over layers $l = 1,2$. Here, $\lambda_l$ and $\mu_l$ are in-plane Lam\'e constants for layer $l$. Summation over repeated indices is implied and we introduced the strain tensor (relative to the rigid moir\'e)
\begin{equation} \label{eq:straintensor}
    u_{ij}^{(l)} = \frac{1}{2} \left( \frac{\partial u_{l,j}}{\partial r_i} + \frac{\partial u_{l,i}}{\partial r_j} \right),
\end{equation}
for $i, j = x, y$. Here, the $\bm u_l$ are the in-plane displacement fields for the acoustic degrees of freedom. To keep the theory analytically tractable, we do not consider out-of-plane deformations at this stage, but we will include them later. Indeed, for heterobilayers, buckling can become important and the quadratic contribution from out-of-plane displacements to the strain tensor can reduce strain at the cost of bending energy \cite{kimE2008}, but may further reduce the elastic energy.

The moir\'e interface between the layers is specified by a rigid displacement gradient \cite{escudero2023}
\begin{equation}
    M = \left( 1 + \frac{\mathcal E}{2} \right) R(\theta/2) - \left( 1 - \frac{\mathcal E}{2} \right) R(-\theta/2),
\end{equation}
where $R$ is a standard counterclockwise rotation matrix and $\mathcal E$ is a symmetric $2\times2$ matrix that encodes both lattice mismatch and external heterostrain. In the long-wavelength limit where the modulation of the interlayer stacking varies slowly compared to the interatomic scale, the moir\'e lattice can be defined by
\begin{equation} \label{eq:moire}
    M \left( \bm r + \bm L \right) = M \bm r + \bm a,
\end{equation}
because a shift by a layer lattice vector $\bm a$ yields the same local stacking configuration. One thus finds
\begin{equation}
    \bm L = M^{-1} \bm a, \qquad \bm g = M^\top \bm b.
\end{equation}
In this work, we only consider rigid matrices $M$ that are invertible. When one eigenvalue of $M$ vanishes, one of the moir\'e lattice vectors diverges, giving a one-dimensional moir\'e \cite{escudero2023}. When both eigenvalues vanish, there is no periodic moir\'e structure. For the case of lattice mismatch (or biaxial strain) and twist, we have
\begin{equation}
    L = |\bm L_1| = |\bm L_2| = \frac{a}{\sqrt{\epsilon^2 \cos^2 \tfrac{\theta}{2} + 4 \sin^2 \tfrac{\theta}{2}}}.
\end{equation}
We further consider adiabatic relaxation such that the displacement fields have the same symmetry as the original rigid moir\'e. Hence, we do not consider broken-symmetry configurations as in Ref.\ \cite{shiPRL2026}, noting that out-of-plane buckling could influence the stability of these states. Hence, for adiabatic relaxation the translational symmetry of the rigid mor\'e is conserved and the in-plane displacement fields can be written as
\begin{equation}
    \bm u_l(\bm r) = \sum_{\bm b} \bm u_{l,\bm b} e^{i \bm b \cdot M \bm r},
\end{equation}
where the sum runs over reciprocal vectors of a fictitious layer that we introduced in Eq.\ \eqref{eq:moire}. In particular, for heterobilayers with lattice mismatch, the lattice constant $a$ of this fictitious layer is given by the mean lattice constant of the physical layers:
\begin{equation}
    a_{1,2} = \left( 1 \pm \frac{\epsilon}{2} \right) a,
\end{equation}
with
\begin{equation}
    a = \frac{a_1 + a_2}{2}, \qquad \epsilon = \frac{a_1 - a_2}{a}.
\end{equation}
For example, for graphene on hBN, taking $a_\text{g} \approx 2.466$~\r A and $a_\text{hBN} \approx 2.513$~\r A obtained from density-functional theory (see Appendix \ref{app:dft} for details) we obtain $a \approx 2.489$~\r A and $\epsilon \approx 0.0187$. Here, we always take $a_1 > a_2$ so that $\epsilon > 0$. The moir\'e lattice constant for $\theta = 0$ (aligned) is then given by $L \approx 14$~nm. The second contribution is the adhesion energy that captures the van der Waals bonding energy between the two layers:
\begin{equation}
    H_\text{adh} = \int d^2\bm r \, \mathcal V \left[ M \bm r + \bm u_1(\bm r) - \bm u_2(\bm r) \right],
\end{equation}
with
\begin{equation}
    \mathcal V(\bm \phi) = \sum_{\bm b} \mathcal V_{\bm b} e^{i\bm b \cdot \bm \phi},
\end{equation}
the stacking potential, shown in Fig.\ \ref{fig:fig1}(c) for a $C_{3v}$ stacking energy landscape corresponding to graphene on hBN or heterobilayers of 2H and 1T TMDs, including aligned \cite{hanNP2026} and anti-aligned \cite{liN2021,zhaoN2023} MoTe$_2$/WSe$_2$ and WS$_2$/WSe$_2$ \cite{gaoNC2024,devenicaNM2026}.

Next, we make use of the fact that a smooth vector field on a torus can always be expressed in a Helmholtz decomposition:
\begin{equation}
    \bm u_{l,\bm b} = \begin{cases} \bm 0 & g = 0, \\ \frac{a}{L} \frac{u_{l,\bm b}^\parallel \bm g + u_{l,\bm b}^\perp \hat z \times \bm g}{i g^2} & g \neq 0, \end{cases} \label{eq.ulb}
\end{equation}
with $g = |\bm g|$ and where the harmonic part (i.e.\ the part that is both divergenceless and irrotational) has to be a constant because the torus has no boundaries. The constant is set to zero, as this only amounts to an overall shift of the moir\'e. Here, the longitudinal ($u^\parallel$) and transverse ($u^\perp$) components of the displacement field give the divergence and curl, respectively. In this way, the elastic energy per unit area becomes
\begin{equation} \label{eq:helas}
    h_\text{elas} = \frac{a^2}{2L^2}  \sum_{l=1,2;\bm b} \left[ \left( \lambda_l + 2 \mu_l \right) | u_{l,\bm b}^\parallel |^2 + \mu_l | u_{l,\bm b}^\perp |^2 \right].
\end{equation}
\begin{table}
    \begin{tabular}{c | c | c | c}
        \Xhline{1pt}
        $n,m$ & $u_{n,m}^\parallel$ & $u_{n,m}^\perp$ & $h_{n,m}$ \\
        \hline
        $1,0$ & $\mathds C$ & $0$ & $\mathds C$ \\ 
        $2,1$ & $\mathds R$ & $i\mathds R$ & $\mathds R$ \\
        $2,0$ & $\mathds C$ & $0$ & $\mathds C$ \\
        $3,1$ & $\mathds C$ & $\mathds C$ & $\mathds C$ \\
        $3,2$ & $(u_{3,1}^\parallel)^*$ & $-(u_{3,1}^\perp)^*$ & $h_{3,1}^*$ \\
        \Xhline{1pt}
    \end{tabular}
    \label{tab:tab1}
    \caption{Constraints on the Fourier components of the adiabatic displacements fields of a heterobilayer moir\'e with $C_{3v} = \left< \mathcal C_{3z}, \mathcal M_x \right>$ symmetry. This corresponds to perfectly aligned or anti-aligned heterobilayers, i.e.\ without any relative twist angle. The reciprocal stars are labeled by the tuple $(n,m)$ with representative $\bm b_{n,m} = n \bm b_1 + m \bm b_2$ where $|\bm b_1| = 4\pi/(\sqrt{3}a)$ and $\bm b_2 = R(2\pi/3) \bm b_1$. Here $n=1,2,\ldots$ is the shell index and $m = 0, \ldots, n-1$ are stars in a given shell.}
\end{table}

\subsection{Self-consistent solution}

Minimizing the total energy with respect to the longitudinal and transverse components, $u_{l,-\bm b}^\parallel$ and $u_{l,-\bm b}^\perp$, respectively, yields \cite{namPRB2017}
\begin{align}
    u_{l,\bm b}^\parallel & = \frac{L}{a} \frac{(-1)^{l+1}}{\lambda_l + 2 \mu_l} \frac{\bm g}{ig^2} \cdot \left( \frac{\partial \mathcal V}{\partial \bm \phi} \right)_{\bm g}, \label{eq:sc1} \\
    u_{l,\bm b}^\perp & = \frac{L}{a} \frac{(-1)^{l+1}}{\mu_l} \frac{\hat z \times \bm g}{ig^2} \cdot \left( \frac{\partial \mathcal V}{\partial \bm \phi} \right)_{\bm g}, \label{eq:sc2}
\end{align}
for $\bm \phi = M \bm r + \bm u_1 - \bm u_2$ and $l = 1, 2$. We immediately notice 
\begin{equation}
    u_{2,\bm b}^\parallel = -\frac{\lambda_1 + 2\mu_1}{\lambda_2 + 2\mu_2} \, u_{1,\bm b}^\parallel, \qquad u_{2,\bm b}^\perp = -\frac{\mu_1}{\mu_2} \, u_{1,\bm b}^\perp,
\end{equation}
such that $(\lambda_1 + 2 \mu_1) \nabla \cdot \bm u_1 = - (\lambda_2 + 2 \mu_2) \nabla \cdot \bm u_2$ and $\mu_1 \nabla \times \bm u_1 = -\mu_2 \nabla \times \bm u_2$ for free-standing heterobilayers, which reduces to the homobilayer result $\bm u_1 = -\bm u_2$ \cite{carrPRB2018,ezziPRL2024,debeulea2025} when $\lambda_1 = \lambda_2$ and $\mu_1 = \mu_2$.

Equations \eqref{eq:sc1} and \eqref{eq:sc2} yield four self-consistent equations for every reciprocal vector $\bm b$ that are solved numerically with the DIIS algorithm \cite{pulayJCC1982}. Generalization to multiple layers is straightforward as long as one can define a single moir\'e lattice with adhesion potentials $\mathcal V_{l,l+1}$ and stacking configurations $\bm \phi_{l,l+1}$. 

\subsection{One-shot result}

To obtain approximate expressions for the relaxed configuration, we solve the self-consistent equations perturbatively in lowest order. This is equivalent to the one-shot result obtained by setting $\bm u_l = \bm 0$ on the right-hand side of Eqs.\ \eqref{eq:sc1} and \eqref{eq:sc2}. We find
\begin{equation}
    \left. \left( \frac{\partial \mathcal V}{\partial \bm \phi} \right)_{\bm g} \right|_\text{one shot} = i \bm b \mathcal V_{\bm b},
\end{equation}
with $\bm g = M^\top \bm b$. Here, the small parameter is the ratio of the characteristic adhesion and elastic energy. For example, $(\mathcal V_\text{max} - \mathcal V_\text{min}) / [(\lambda + 2 \mu) \epsilon^2]$ for $\theta = 0$. Hence, the one-shot solution is given by
\begin{align}
    u_{l,\bm b}^\parallel & \simeq \frac{L}{a} \frac{(-1)^{l+1}}{\lambda_l + 2 \mu_l} \frac{\bm b \cdot M \bm b}{g^2} \, \mathcal V_{\bm b}, \\
    u_{l,\bm b}^\perp & \simeq \frac{L}{a} \frac{(-1)^{l+1}}{\mu_l} \frac{\left( \hat z \times \bm b \right) \cdot M \bm b}{g^2} \, \mathcal V_{\bm b},
\end{align}
which is applicable to any moir\'e bilayer composed of layers with the same Bravais lattice type. For lattice mismatch $\epsilon$ and twist $\theta$, we obtain
\begin{align}
    u_{l,\bm b}^\parallel & \simeq \frac{(-1)^{l+1}}{\lambda_l + 2 \mu_l} \frac{\epsilon \mathcal V_{\bm b} \cos \frac{\theta}{2}}{\left( \epsilon^2 \cos^2 \frac{\theta}{2} + 4 \sin^2 \frac{\theta}{2} \right)^{3/2}}, \\
    u_{l,\bm b}^\perp & \simeq \frac{(-1)^{l+1}}{\mu_l} \frac{2 \mathcal V_{\bm b} \sin \frac{\theta}{2}}{\left( \epsilon^2 \cos^2 \frac{\theta}{2} + 4 \sin^2 \frac{\theta}{2} \right)^{3/2}}.
\end{align}

We now take a $C_{3v}$ stacking energy:
\begin{equation} \label{eq:stacking}
    \mathcal V(\bm \phi) = 2 \mathcal V_1 \sum_{n=1}^3 \cos\left( \bm b_n \cdot \bm \phi + \varphi_1 \right),
\end{equation}
where we take $\mathcal V_1 > 0$ and which is a first-star approximation for graphene on hBN and 2H TMD heterobilayers with $\mathcal V_{\bm b_1} = \mathcal V_{\bm b_2} = \mathcal V_{\bm b_3} = \mathcal V_1 e^{i\varphi_1}$. Here, we take reciprocal vectors $\bm b_1 = 4\pi \hat y/(\sqrt{3}a)$ and $\bm b_{n+1} = R(2\pi/3) \bm b_n$. For graphene on hBN, we use the parameters listed in Tables \ref{tab:tab2} and \ref{tab:tab3} obtained from density-functional theory calculations for untwisted bilayers (see Appendix \ref{app:dft}).
\begin{table}
    \centering
    \begin{tabular}{l | c | c | c | c}
        \Xhline{1pt}
        & $a$ & $\lambda$ & $\mu$ & $\nu$ \\
        \hline
        graphene & $2.466$ & $26756$ & $44702$ & $0.230$ \\
        \hline
        hBN & $2.513$ & $22621$ & $37791$ & $0.230$ \\
        \hline
        2H WS$_2$ & $3.188$ & $30226$ & $31405$ & $0.323$ \\
        \hline
        2H WSe$_2$ & $3.322$ & $24587$ & $30099$ & $0.290$ \\
        \hline
        2H MoTe$_2$ & $3.567$ & $22130$ & $24161$ & $0.314$ \\
        \Xhline{1pt}
    \end{tabular}
    \label{tab:tab2}
    \caption{Lattice constant $a$ in \r A, 2D Lam\'e parameters $\lambda$ and $\mu$ in units of meV per monolayer unit cell, and the 2D Poisson's ratio $\nu = 1 / \left( 1 + 2 \mu / \lambda \right)$.}
\end{table}

In general, the acoustic displacement fields of the adiabatic relaxed ground-state configuration are constrained by the symmetry of the rigid moir\'e \cite{ezziPRL2024}. This results in symmetry constraints for the Fourier components of each reciprocal star, given in Table \ref{tab:tab1}. Here, a star is defined as a set of six reciprocal vectors closed under $\mathcal C_{6z}$. The self-consistent solution inherits these symmetries from the adhesion potential. In lowest order,
\begin{equation} \label{eq:uloneshot}
    \begin{aligned}
        \bm u_l(\bm r) \simeq \frac{\sqrt{3}a}{2\pi} \sum_{n=1}^3 & \left( u_{l,1}^\parallel \hat g_n  + u_{l,1}^\perp \hat z \times \hat g_n \right) \\
        & \times \sin( \bm g_n \cdot \bm r + \varphi_1 ),
    \end{aligned}
\end{equation}
with the one-shot result
\begin{align}
    u_{l,m}^\parallel & = \frac{(-1)^{l+1}}{\lambda_l + 2 \mu_l} \frac{\epsilon \mathcal V_m}{( \epsilon^2 + \theta^2 )^{3/2}}, \\
    u_{l,m}^\perp & = \frac{(-1)^{l+1}}{\mu_l} \frac{\theta \mathcal V_m}{( \epsilon^2 + \theta^2 )^{3/2}},
\end{align}
for small $\theta$ and where $m$ labels the star. These expressions recover the results from Refs.\ \cite{ezziPRL2024,ceferino2M2024} for homobilayers with $\epsilon \rightarrow 0$ and finite $\theta$. 
\begin{figure}
    \centering
    \includegraphics[width=\linewidth]{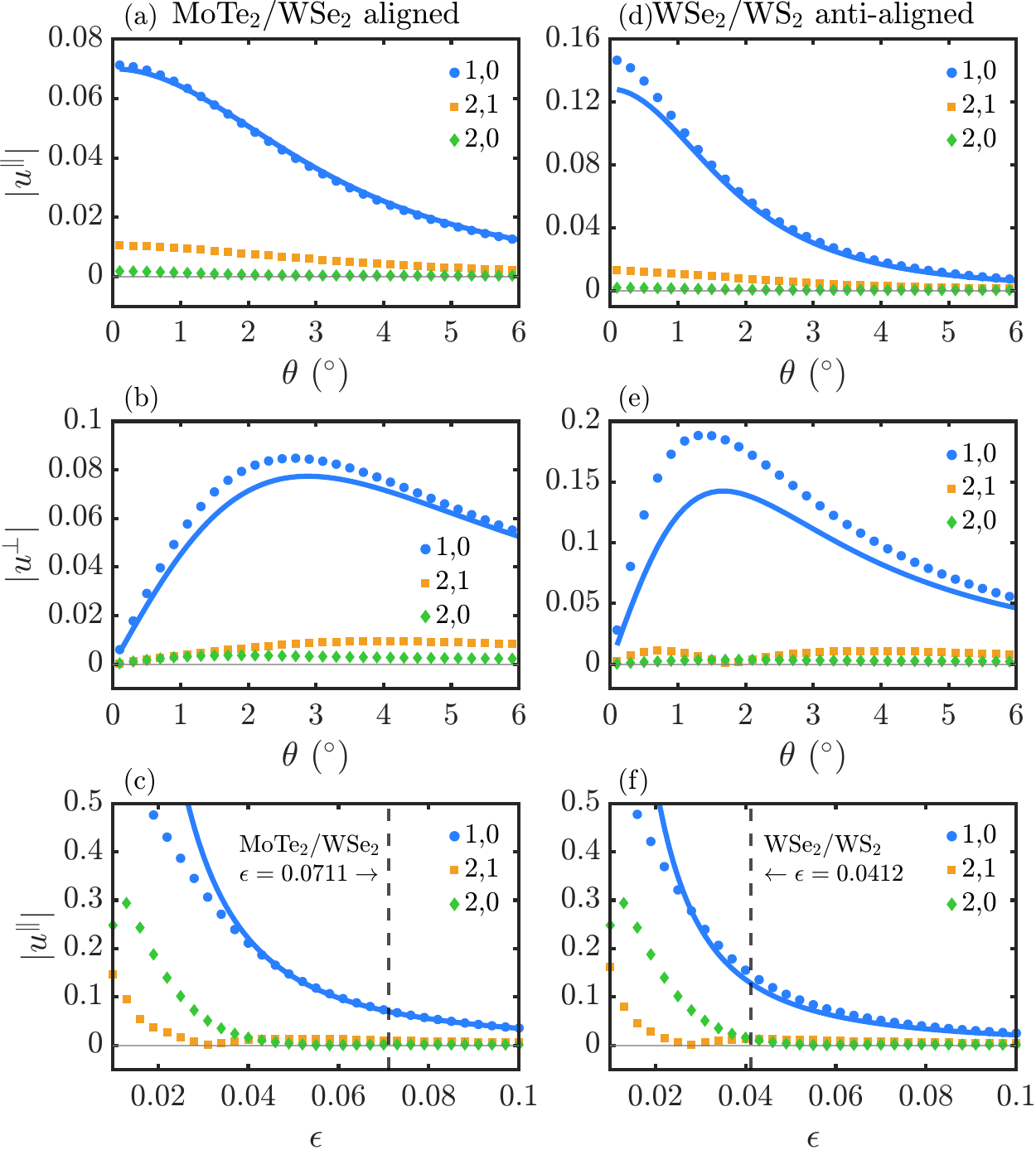}
    \caption{Acoustic displacement fields induced by atomic reconstruction in moir\'e TMD heterobilayers. Here, we show the longitudinal and transverse Fourier components of the first reciprocal star for the in-plane displacement fields $\bm u = \bm u_1 - \bm u_2$ where the dots are numerical solutions of the self-consistent equations given in Eqs.\ \ref{eq:sc1} and \ref{eq:sc2}, and the curves give the one-shot analytical result. Shown as a function of the twist angle for aligned MoTe$_2$/WSe$_2$ in panels (a) and (b), and for anti-aligned WSe$_2$/WS$_2$ in panels (d) and (e). We also show the transverse component versus $\epsilon$ for aligned MoTe$_2$/WSe$_2$ in panels (c) and for anti-aligned WSe$_2$/WS$_2$ in panels (f) for $\theta = 0$. The center-of-mass fields are obtained from $U_{\bm b}^\parallel = A^\parallel u_{\bm b}^\parallel$ and $U_{\bm b}^\perp = A^\perp u_{\bm b}^\perp$. For aligned MoTe$_2$/WSe$_2$, $A^\parallel \approx 0.09$ and $A^\perp \approx 0.11$. For anti-aligned WSe$_2$/WS$_2$, $A^\parallel \approx 0.05$ and $A^\perp \approx 0.02$. Parameters are shown in Tables \ref{tab:tab2} and \ref{tab:tab3}.}
    \label{fig:fig2}
\end{figure}

It is now instructive to write the original layer-resolved fields in terms of the layer center-of-mass ($\bm U$) and relative displacements ($\bm u$). This is motivated by the fact that the latter is dominant, while the former is expected to be much smaller. We define these auxiliary fields as
\begin{equation}
    \bm u_{1,2} = \frac{\bm U \pm \bm u}{2}.
\end{equation}
Moreover,
\begin{align}
    U_{\bm b}^\parallel & = \frac{\lambda_2 + 2 \mu_2 - \lambda_1 - 2 \mu_1}{\lambda_2 + 2 \mu_2 + \lambda_1 + 2 \mu_1} \, u_{\bm b}^\parallel \equiv A^\parallel u_{\bm b}^\parallel, \\
    U_{\bm b}^\perp & = \frac{\mu_2 - \mu_1}{\mu_2 + \mu_1} \, u_{\bm b}^\perp \equiv A^\perp u_{\bm b}^\perp,
\end{align}
such that we only specify $u_{\bm b}^\parallel$ and $u_{\bm b}^\perp$. This works because the longitudinal and transverse components are decoupled in the elastic energy. It further implies $\bm U = A^\parallel \nabla \phi + A^\perp \nabla \times \bm w$, with $\bm u = \nabla \phi + \nabla \times \bm w$.

In Appendix \ref{app:sc} we also show the elastic energy in terms of these fields. We compare these perturbative results to the numerical results in Fig.\ \ref{fig:fig1}. We find excellent agreement between the analytical theory and the numerics for the case of graphene on hBN. We see that the longitudinal components (related to local compression and dilation) are largest in magnitude for $\theta_\parallel = 0$ while the transverse components (related to twirling or local rotations) are maximal for $\theta_\perp = \epsilon / \sqrt{2}$. For graphene on hBN, we find $\theta_\perp \approx 0.76^\circ$. In addition, we also apply the analytical theory to several 2H TMD heterobilayers in Fig.\ \ref{fig:fig2}. Now the stacking energy contains non-negligible contributions from the second and third star. For example, up to the second star, $\bm u_l  \simeq \bm u_{l,1} + \bm u_{l,2}$ with
\begin{equation}
    \bm u_{l,2}(\bm r) = \frac{a}{2\pi} \sum_{n=1}^3 \left( u_{l,2}^\parallel \hat g_n'  + u_{l,2}^\perp \hat z \times \hat g_n' \right) \sin(\bm g_n' \cdot \bm r),
\end{equation}
where $\bm g_1' = 2\bm g_1 + \bm g_2$ and $\bm g_{2,3}'$ related by $120^\circ$ rotations, are reciprocal vectors of the second star. Thus, the relaxed stacking configuration in experimentally relevant heterobilayers (aligned, anti-aligned, or twisted) can be accurately captured using a perturbative solution. This is because the moir\'e length in these systems is bounded from above by the lattice mismatch.
\begin{table*}
    \centering
    \begin{tabular}{l | c | c | c | c | c | c | c | c | c}
        \Xhline{1pt}
        layer 1/layer 2 & $\epsilon$ & $a$ & $\mathcal V_1$ & $\mathcal V_2$ & $\mathcal V_3$ & $\mathcal V_{4,5}$ & $\varphi_1$ & $\varphi_3$ & $\varphi_4 = -\varphi_5$ \\
        \hline
        hBN/graphene & $0.0187$ & $2.489$ & $1.421$ & $-0.07514$ & $0.01734$ & $0.02080$ & $0.8867$ & $-1.491$ & $2.535$ \\
        \hline
        P WSe$_2$/WS$_2$ & $0.0412$ & $3.255$ & $11.09$ & $-2.090$ & $0.7501$ & $0.2748$ & $0.06173$ & $-2.958$ & $-0.07944$ \\
        \hline
        P MoTe$_2$/WSe$_2$ & $0.0711$ & $3.445$ & $13.61$ & $-2.675$ & $0.9122$ & $0.3566$ & $0.04351$ & $-2.985$ & $-0.07550$ \\
        \hline
        AP WSe$_2$/WS$_2$ & $0.0412$ & $3.255$ & $9.665$ & $-1.608$ & $0.6338$ & $0.1922$ & $-2.301$ & $-1.375$ & $2.144$ \\
        \hline
        AP MoTe$_2$/WSe$_2$ & $0.0711$ & $3.445$ & $12.04$ & $-2.079$ & $0.8117$ & $0.2598$ & $2.354$ & $1.480$ & $-2.161$ \\     
        \Xhline{1pt}
    \end{tabular}
    \label{tab:tab3}
    \caption{Lattice mismatch $\epsilon$, average lattice constant $a$ in units of \r A, and Fourier coefficients of the adhesion potential in units of meV per monolayer unit cell (taking the average lattice constant) for the heterobilayers considered in this work. Angles are given in radians. We always take $a_1 > a_2$ such that $\epsilon > 0$ and we consider both aligned (P) and anti-aligned (AP) stacking.} 
\end{table*}

\section{Buckling}

We now consider the out-of-plane displacement fields $h_l$ ($l=1,2$). Here, we distinguish between breathing $h_1 - h_2$ and buckling $h_1 + h_2$. Buckling can be driven by in-plane compression to reduce strain \cite{baoNN2009,caiJotMaPoS2011,cerdaPRL2003,maoN2020,wangPNAS2024}. Similarly, substrate engineering of out-of-plane deformations in graphene gives rise to local in-plane compressive deformations \cite{phongPRL2022,debeulePNAS2023,debeulePRL2025}. In free-standing twisted homobilayers, compression-driven buckling is absent as $\nabla \cdot \bm u_l \approx 0$ \cite{ezziPRL2024,debeulea2025}. Instead, homobilayers may exhibit shear-driven buckling with spontaneously broken moir\'e translational symmetry \cite{wangPNAS2024}. Here, we consider buckling in heterobilayers without symmetry breaking. We thus expect that the buckling amplitude is largest for aligned (or anti-aligned) layers, for which the longitudinal components (and thus local compression) are dominant. With increasing twist angle, the latter and thus buckling are suppressed until a buckling transition occurs at a critical twist angle.

Here, we present a simplified theory of buckling by considering a single layer in the presence of the relaxation-induced in-plane displacement fields that we obtained in the previous section. We then minimize the elastic energy of this layer with respect to the height profile. This theory captures the qualitative features of the buckling transition, while details of the buckling profile and the critical angle are expected to differ in a more accurate theory. Such a theory should include the coupling between buckling and breathing modes through the interlayer interaction, i.e.\ one needs to include the dependence of the interlayer separation in the Fourier components of the adhesion potential \cite{enaldievPRL2020}. 

\subsection{Symmetry ansatz}

The details of the buckling theory are presented in Appendix \ref{app:buckling}. To simplify the calculation, we use an \textit{ansatz} for the height profile with $C_{3v}$ symmetry:
\begin{equation} \label{eq:hr}
    \begin{aligned}
        h(\bm r) & = 2 \sum_{n=1}^3 \Big[ h_1 \cos(\bm g_n \cdot \bm r + \gamma_1) + h_2 \cos(\bm g_n' \cdot \bm r) \\
        & + h_3 \cos(2\bm g_n \cdot \bm r + \gamma_3) + \left( \text{higher stars} \right) \Big],
    \end{aligned}
\end{equation}
with amplitudes $h_n$ and phases $\varphi_n$. The elastic energy is then computed analytically and numerically minimized with respect to the amplitudes and phases. We have to go beyond the first star because the height profile enters nonlinear in the strain. This is similar to the reverse problem, where for a fixed first-star buckling profile, the elastic ground state contains up to three stars of the in-plane field \cite{debeulePRL2025}. We find that the in-plane rotational components ($u^\perp$) do not couple to the height profile at this order of approximation. This is enforced by symmetry as detailed in Appendix \ref{app:buckling}. Moreover, the transverse in-plane components are also decoupled from the longitudinal ones ($u^\parallel$) [see Eq.\ \eqref{eq:helas}], such that the $u^\perp$ do not affect buckling.

The in-plane displacement field due to structural relaxation in moir\'e heterobilayers gives rise to both regions of in-plane tensile and compressive strain. Because the volumetric part of the strain tensor from the height modulation is given by $(1/2)(\nabla h)^2 \geq 0$, regions of in-plane tensile strain favor a constant height profile, while compressive regions favor a varying $h(\bm r)$ to lower the total strain. These gains compete with the bending energy $( \kappa / 2) (\nabla^2 h)^2$ which disfavors out-of-plane deformations.
\begin{figure}
    \centering
   \includegraphics[width=\linewidth]{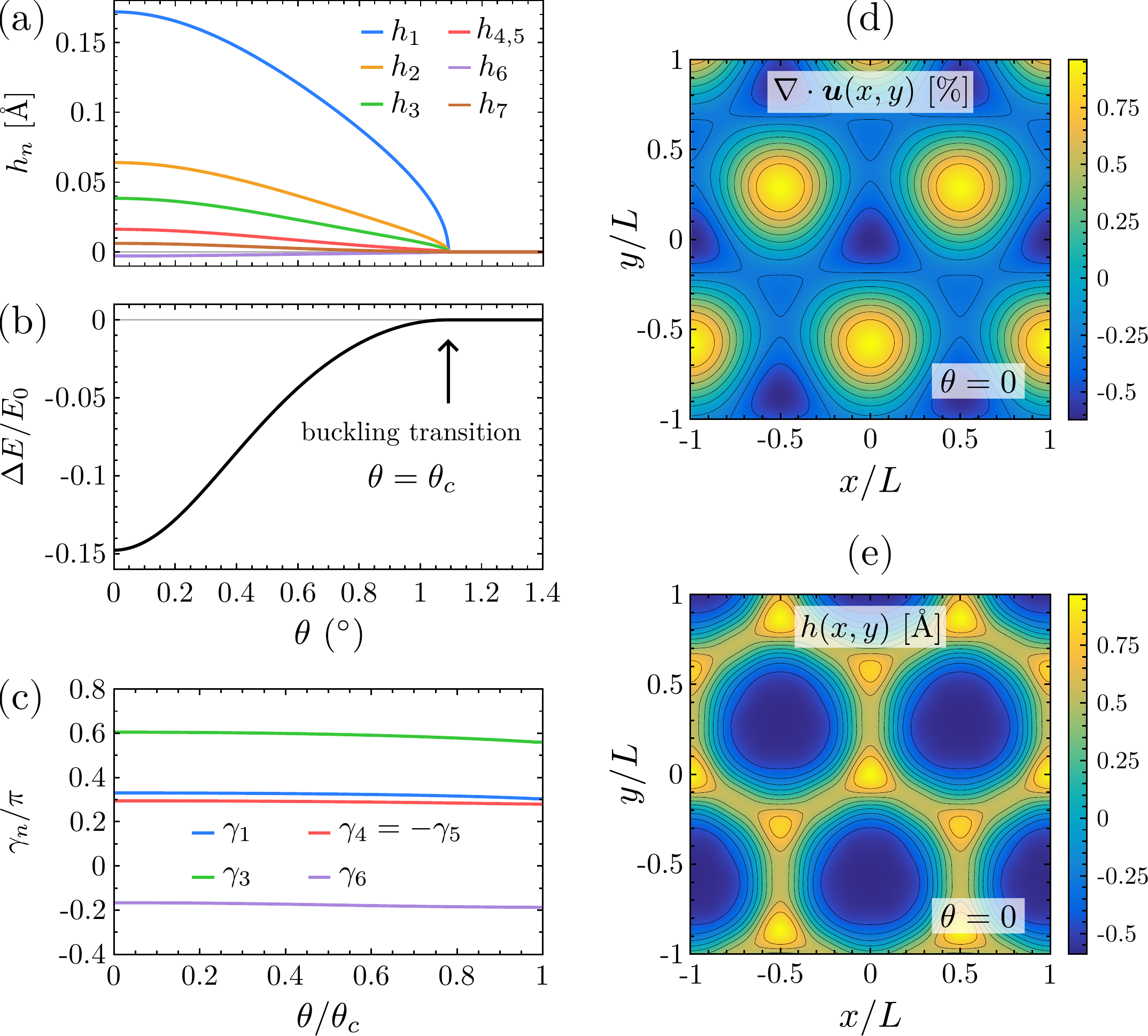}
    \caption{Buckling transition in moir\'e heterobilayers. Shown for graphene on hBN for $\kappa = 1$~eV and the parameters in Tables \ref{tab:tab2} and \ref{tab:tab3}. (a) Fourier amplitudes $h_n$ as a function of twist angle for a $C_{3v}$ buckling profile with up to 7 stars. (b) Relative energy gain due to buckling versus twist angle. (c) Phases of the Fourier components. The 2nd and 7th star are real by symmetry and not shown. (d) Divergence of the in-plane displacement field for graphene aligned to hBN ($\theta = 0$) using the one-shot analytical result. (e) Buckling profile corresponding to (d) using the results from (a) and (c).}
    \label{fig:fig3}
\end{figure}

\subsection{Buckling transition}

We start by considering only the first star in the \textit{ansatz} from Eq.\ \eqref{eq:hr}. While, we numerically find that we need up to seven stars for convergence, we can analytically demonstrate that a buckling transition can occur in principle. In particular, we find
\begin{align}
    h_1^2 = \frac{(-1)^n 3a L u_1 \left( \mu - \lambda \right) / \pi^2 - 16 \kappa}{72 \left( \lambda + 2 \mu \right)}, \label{eq:buckling}
\end{align}
with $\gamma_1 = -\left( \varphi_1 + n\pi \right) / 2$ and $n = 0, 1$. Here, $a = L \sqrt{\epsilon^2 + \theta^2}$ and $u_1$ is the longitudinal first-star component. Since $h_1$ is real, this requires
\begin{equation}
    (-1)^n u_1 > \frac{\kappa}{\mu - \lambda} \frac{16\pi^2}{3aL},
\end{equation}
or $h_1 = 0$ otherwise. Because we consider a single layer, the solution is degenerate under $h_1 \rightarrow -h_1$. In a moir\'e bilayer, there is a preferential orientation due to coupling between the buckling and breathing modes. Using our previous result: $u_1 = (-1)^{l+1} \mathcal V_1 \epsilon / [(\lambda+2\mu)(\epsilon^2 + \theta^2)^{3/2}]$ for layer $l=1,2$, together with Eq.\ \eqref{eq:buckling}, we find a nontrivial buckling solution for
\begin{equation}
    \frac{\epsilon}{\left( \epsilon^2 + \theta^2 \right)^2} > \frac{32 \pi^2}{1 - 3 \nu} \frac{\kappa}{3 a^2 \mathcal V_1},
\end{equation}
where $n = 0$ for the first layer and $n = 1$ for the second layer with $a_1 > a_2$. Here, we defined $\nu = 1 / \left( 1 + 2 \mu / \lambda \right)$ the 2D Poisson's ratio and $a$ the average lattice constant of the two layers. Thus, the first-star approximation predicts a buckling transition for $\epsilon \sim \kappa a^2 / ( \mathcal V_1 L^4 )$ or below a critical twist angle
\begin{equation}
    \left. \theta_c^2 \right|_\text{first star} = \frac{a}{4\pi} \sqrt{\frac{3\epsilon (1-3\nu) \mathcal V_1}{2\kappa}} - \epsilon^2,
\end{equation}
determined by the ratio $\epsilon (1 - 3\nu) a^2 \mathcal V_1 / \kappa$. This is to be expected, as the energy cost for buckling is proportional to $\kappa$ while the gain is proportional to $|u_1| \propto \mathcal V_1$.  While this result provides good intuition, an accurate determination of $\theta_c$ is done numerically. We take a $C_{3v}$ buckling profile up to the seventh star, and numerically compute the coefficients $h_{\bm g}$ that minimize the elastic energy for a fixed in-plane displacement field from Eq.\ \eqref{eq:uloneshot}. Here, we only consider the layer with the smaller lattice constant. Note that the in-plane displacement field for the other layer has the opposite sign, and the resulting buckling profiles are not equivalent. Using $\kappa = 1$~eV, we find a buckling transition for graphene on hBN below critical twist angle $\theta_c \approx 1.1^\circ$. This is illustrated in Fig.\ \ref{fig:fig3}(a) and (b) where we show the $h_n$ for each star and the relative energy gain, respectively, as a function of twist angle. We also show the corresponding phases $\gamma_n$ in Fig.\ \ref{fig:fig3}(c). The in-plane volumetric strain and buckling profile for $\theta = 0$ are shown in Fig.\ \ref{fig:fig3}(d) and (e), respectively. We see that compressive in-plane strain ($\nabla \cdot \bm u < 0$) favors buckling, while tensile strain ($\nabla \cdot \bm u > 0$) disfavors buckling, leading to a bucket-shaped buckling profile. For aligned (anti-aligned) WSe$_2$/WS$_2$ we find $\theta_c \approx 2.2^\circ (2^\circ)$, while we find no buckling for MoTe$_2$/WSe$_2$ heterobilayers.

\section{Conclusion}

In this work, we developed an analytical theory of lattice relaxation in moir\'e heterobilayers. We modeled the intralayer elastic properties with continuum elasticity and the interlayer energy with the local stacking approximation using parameters obtained from density functional theory for untwisted bilayers. We then derived the self-consistent equations for the Fourier components of the acoustic in-plane displacement fields and obtained the one-shot perturbative result. We subsequently applied this theory to graphene on hBN and representative 2H transition metal dichalcogenide heterobilayers, and found good agreement between the analytical results and the full numerical solutions over experimentally relevant parameters. These results show that the relaxed stacking configuration in realistic heterobilayer moir\'e systems can be captured accurately within a simple analytical framework, even in the presence of both lattice mismatch and twist. In addition, using our analytical result for the in-plane displacement field, we demonstrated that moir\'e heterobilayers exhibit a buckling instability near alignment, driven by compressive strain due to in-plane relaxation. More broadly, our results provide both a practical framework for treating lattice relaxation in realistic moir\'e heterostructures and a foundation for integrating structural reconstruction into calculations of their electronic and optical properties.

\begin{acknowledgments}
This work was supported by a start-up grant from Washington University in St. Louis. C.D.B.\ acknowledges financial support from the Methusalem funding of the University of Antwerp. D.B.\ acknowledges support from the NTU Startup Grant (Award Number 025661-00003). The \textsc{Julia} code used to solve the continuum elasticity equations of motion self-consistently is available on GitHub \cite{debeule2026}.  
\end{acknowledgments}

\appendix

\section{First-principles calculations} \label{app:dft}

First-principles density functional theory (DFT) calculations were performed to simulate graphene, hBN, and TMDs using the {\sc abinit}~\cite{gonze2009abinit,gonze2020,verstraete2025abinit} code.
PAW pseudopotentials were used~\cite{torrent2008implementation}, obtained from Pseudo-Dojo~\cite{pseudodojo}, and the PBE functional was used~\cite{pbe} to treat the exchange-correlation interactions.
{\sc abinit} employs a plane-wave basis set, which was determined using a kinetic energy cutoff of 1200 eV, and a cutoff of 1500 eV was used for PAW double grid.
A Monkhorst-Pack $k$-point grid~\cite{mp} of $12 \times 12 \times 1$ was used to sample the Brillouin zone for hBN and the TMDs, and a sampling of $21 \times 21 \times 1$ was used for graphene, which requires a finer sampling as it is not gapped.

Geometry relaxations were performed for the graphene, hBN, WS$_2$, WSe$_2$ and MoTe$_2$ monolayers to obtain the equilibrium lattice constants and elastic properties.
The bulk and shear moduli were calculated by making small deviations of the lattice constant and the angle between the lattice vectors, respectively, and taking the second derivatives of the change in energy. The equilibrium lattice constants, Lam\'e factors, and Poisson ratios are given in Table \ref{tab:tab2}.

Commensurate hBN/graphene, WS$_2$/WSe$_2$, and MoTe$_2$/WSe$_2$ heterostructures were then constructued, using the average of the lattice constants of the individual monolayers in each case.
To sample the different local environments, a grid containing $31$ stacking configurations not related by symmetry on a triangular grid was used, which is compatible with the $C_{3v}$ symmetry of the adhesion potential of the heterostructures.
For each relative stacking, a geometry relaxation was performed to obtain the equilibrium layer separation, while keeping the in-plane lattice vectors and atomic positions fixed, using a force tolerance of 1 meV/Å. The energy as a function of stacking was then linearly interpolated and numerically Fourier transformed to obtain the coefficients
\begin{equation}
    \mathcal V_{\bm b} = \frac{1}{A_c} \int_\text{cell} d^2\bm r \, \mathcal V(\bm r) e^{-i\bm b \cdot \bm r},
\end{equation}
given in Table \ref{tab:tab2}.

\section{Self-consistent equations} \label{app:sc}

\begin{figure}
    \centering
    \includegraphics[width=0.9\linewidth]{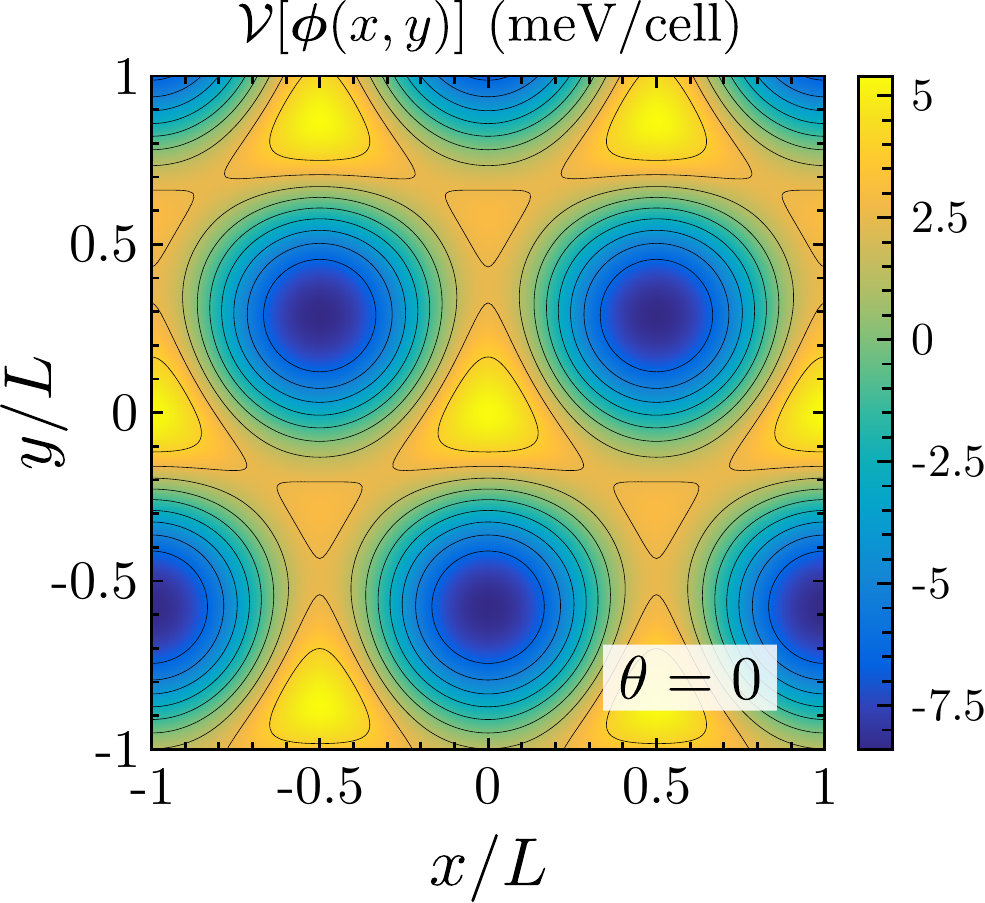}
    \caption{Stacking energy landscape (up to an overall constant) in units of meV per monolayer unit cell (using the average lattice constant). Shown for the relaxed moir\'e configuration $\bm \phi(\bm r) = \epsilon \bm r + \bm u(\bm r)$ for graphene aligned on hBN. Here, we used the analytical result for the relaxed in-plane acoustic displacement fields. Parameters used are listed in Tables \ref{tab:tab2} and \ref{tab:tab3} of the main text.}
    \label{fig:app0}
\end{figure}

\begin{figure*}
    \centering
    \includegraphics[width=\linewidth]{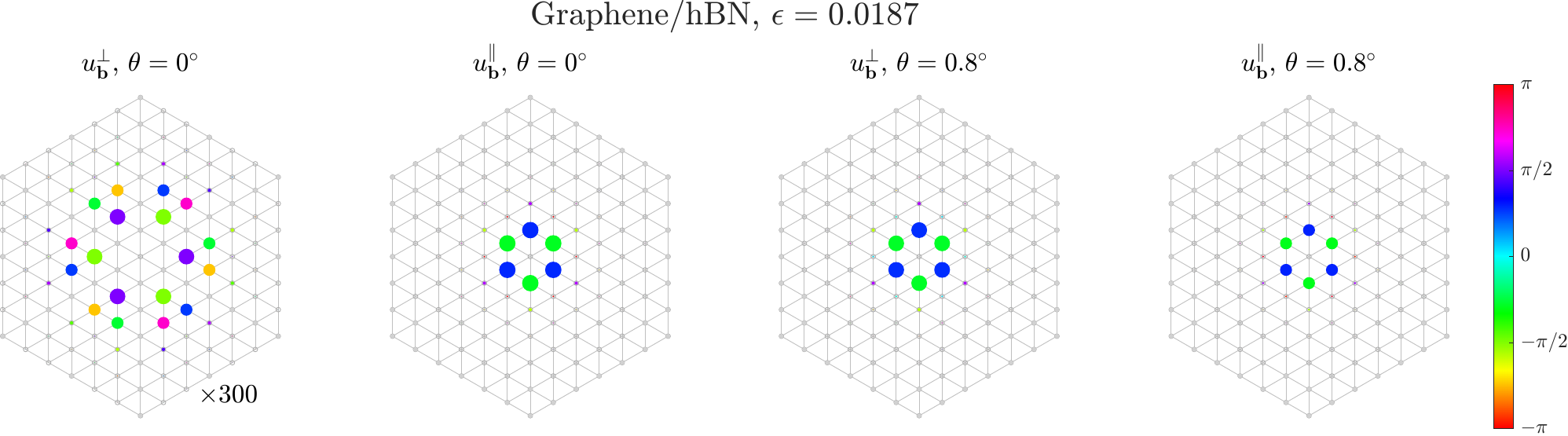}
    \caption{    Fourier components of the relative displacement field
    \(\mathbf u=\mathbf u_1-\mathbf u_2\) for graphene on hBN at
    \(\theta=0^\circ\) and \(\theta=0.8^\circ\), with \(\epsilon=0.0187\).
    The transverse and longitudinal components, \(u^\perp_{\mathbf b}\) and \(u^\parallel_{\mathbf b}\), are shown up to the sixth shell. Dot size denotes \(|u_{\mathbf b}|\), and color denotes the phase
    \(\arg(u_{\mathbf b})\). The indicated scale factors are applied to the dot sizes for visual clarity.}
    \label{fig:app1}
\end{figure*}

\begin{figure*}
    \centering
    \includegraphics[width=\linewidth]{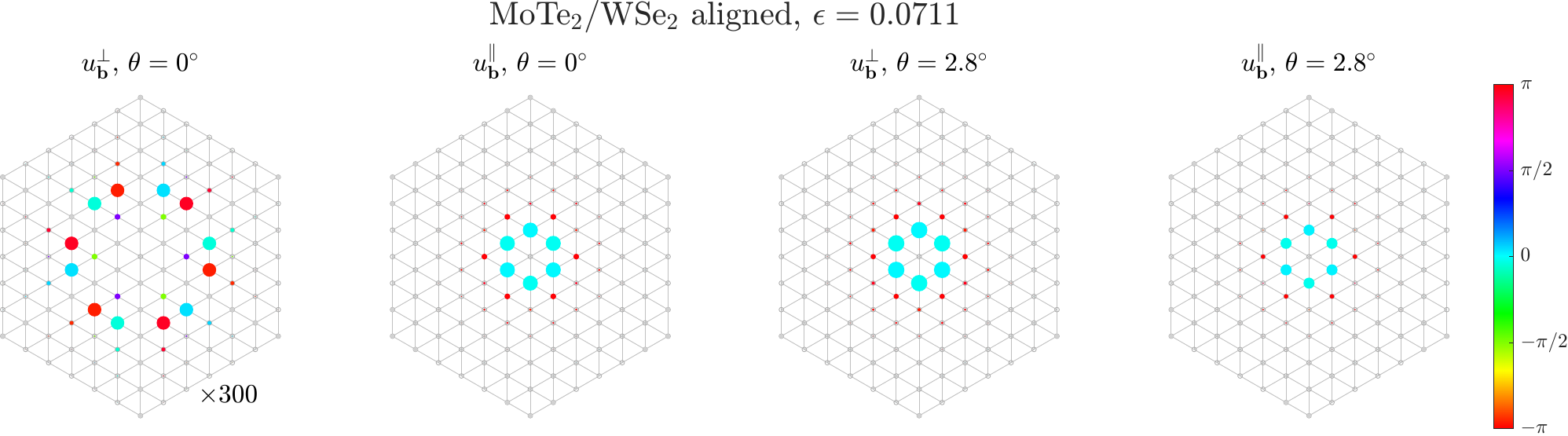}
    \vspace{1em}
    \includegraphics[width=\linewidth]{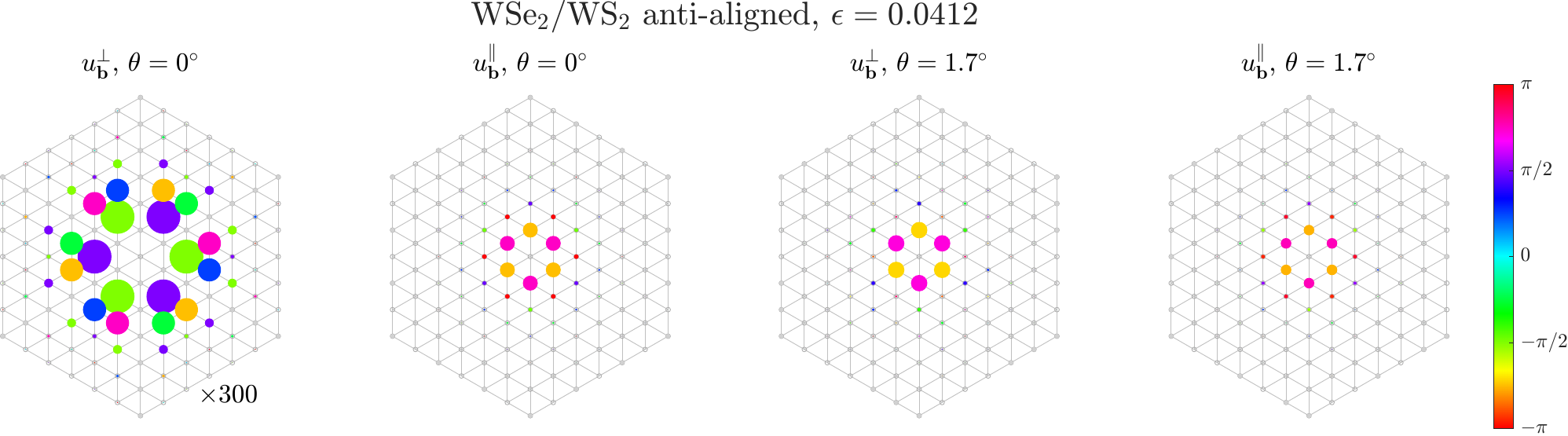}
    \caption{Fourier components of the relative displacement field
    \(\mathbf u=\mathbf u_1-\mathbf u_2\) for TMD heterobilayers.
    The top row shows aligned \(\mathrm{MoTe}_2/\mathrm{WSe}_2\) with
    \(\epsilon=0.0711\), while the bottom row shows anti-aligned
    \(\mathrm{WSe}_2/\mathrm{WS}_2\) with \(\epsilon=0.0412\). For each
    system, \(u^\perp_{\mathbf b}\) and \(u^\parallel_{\mathbf b}\) are shown
    at \(\theta=0^\circ\) and at the representative finite twist angle labeled
    above each panel. Results are shown up to the sixth shell. Dot size denotes
    \(|u_{\mathbf b}|\), color denotes \(\arg(u_{\mathbf b})\), and the
    indicated scale factors are used for visual clarity.}
    \label{fig:app2}
\end{figure*}

In this section, we give the full derivation of the self-consistent equations obtained from continuum elasticity, for the in-plane displacement field of a general moir\'e bilayer. To this end, we first write down the Fourier transform of the strain tensor:
\begin{equation}
    u_{ij} = \frac{i}{2} \sum_{\bm b} \left( g_i u_{l,\bm b,j} + g_j u_{l,\bm b,i} \right) e^{i \bm g \cdot \bm r},
\end{equation}
where $\bm g = M^\top \bm b$ so we can freely move between the atomic lattice scale $\bm b$ and the moir\'e scale $\bm g$. Plugging this into the expression for the elastic energy and integrating, using
\begin{equation}
    \frac{1}{A} \int d^2\bm r \, f(\bm r) e^{-i\bm g \cdot \bm r} = f_{\bm g},
\end{equation}
with $A$ the total area and $f(\bm r) = f(\bm r + \bm L)$ a periodic function, we find
\begin{widetext}
\begin{align}
    h_\text{elas} & = \frac{1}{2} \sum_{l=1,2} \sum_{\bm b} \left[ \lambda_l \left| \bm g \cdot \bm u_{l,\bm b} \right|^2 + \frac{\mu_l}{2} \left| g_i u_{l,\bm b,j} + g_j u_{l,\bm b,i} \right|^2 \right] \\
    & = \frac{1}{2} \sum_{l=1,1} \sum_{\bm b} \left[ \left( \lambda_l + 2 \mu_l \right) \left| \bm g \cdot \bm u_{l,\bm b} \right|^2 + \mu_l \left| \left( \hat z \times \bm g \right) \cdot \bm u_{l,\bm b} \right|^2 \right] \\
    & = \frac{a^2}{2L^2} \sum_{l=1,1} \sum_{\bm b} \left[ \left( \lambda_l + 2 \mu_l \right) | u_{l,\bm b}^\parallel |^2 + \mu_l | u_{l,\bm b}^\perp |^2 \right],
\end{align}
where we used that a smooth vector field on a torus can always be expressed in a Helmholtz decomposition:
\begin{equation}
    \bm u_{l,\bm b} = \begin{cases} \bm 0 & g = 0, \\ \frac{a}{L} \frac{u_{l,\bm b}^\parallel \bm g + u_{l,\bm b}^\perp \hat z \times \bm g}{i g^2} & g \neq 0, \end{cases}
\end{equation}
with $g = |\bm g|$ and where the harmonic part (i.e.\ a gradient $\nabla f$ with $\nabla^2 f = 0$) has to be a constant because the torus has no boundaries. This constant is set to zero, as this only amounts to an overall shift of the moir\'e lattice. The longitudinal and transverse components of the displacement field give the divergence and curl, respectively,
\begin{align}
    \nabla \cdot \bm u_l & = \frac{a}{L} \sum_{\bm h} u_{l,\bm g}^\parallel e^{i\bm g \cdot \bm r}, \\
    \nabla \times \bm u_l & = \frac{a}{L} \sum_{\bm h} u_{l,\bm g}^\perp e^{i\bm g \cdot \bm r}.
\end{align}
We also find
\begin{align}
    h_\text{adh} & = \frac{1}{A_m} \int_\text{moir\'e cell} d^2\bm r \, \mathcal V \left[ M \bm r + \bm u_1(\bm r) - \bm u_2(\bm r) \right] \\
    & = \frac{1}{A_m} \sum_{\bm b} \mathcal V_{\bm b} \int_\text{moir\'e cell} d^2 \bm r \, \exp \left\{ i \bm b \cdot \left[ M \bm r + \bm u_1(\bm r) - \bm u_2(\bm r) \right] \right\},
\end{align}
with $A_m = |\bm L_1 \times \bm L_2| = (L/a)^2 |\bm a_1 \times \bm a_2|$ the moir\'e unit cell area. The ground-state is obtained by minimizing the total elastic and adhesion energy with respect to $\{u_{l,\bm b}^\parallel, u_{l,\bm b}^\perp\}$ for $l = 1,2$. We find
\begin{equation}
    \frac{\partial h_\text{elas}}{\partial u_{l,-\bm b}^\parallel} = \frac{a^2}{L^2} \left( \lambda_l + 2 \mu_l \right) u_{l,\bm b}^\parallel, \qquad \frac{\partial h_\text{elas}}{\partial u_{l,-\bm b}^\perp} = \frac{a^2}{L^2} \mu_l u_{l,\bm b}^\perp,
\end{equation}
and
\begin{align}
    \frac{\partial h_\text{adh}}{\partial u_{l,-\bm b}^\parallel} & = \frac{1}{A_m} \int_\text{moir\'e cell} d^2 \bm r \, \frac{\partial \bm u_l}{\partial u_{l,-\bm b}^\parallel} \cdot \frac{\partial \mathcal V}{\partial \bm u_l} \left[ M \bm r + \bm u_1(\bm r) - \bm u_2(\bm r) \right] \\
    & = -\frac{a}{L} \frac{\bm g}{ig^2} \cdot \frac{1}{A_m} \int_\text{moir\'e cell} d^2 \bm r \, e^{-i\bm g \cdot \bm r} \frac{\partial \mathcal V}{\partial \bm u_l} \left[ M \bm r + \bm u_1(\bm r) - \bm u_2(\bm r) \right] \\
    & = (-1)^l \frac{a}{L} \frac{\bm g}{ig^2} \cdot \left( \frac{\partial \mathcal V}{\partial \bm \phi} \right)_{\bm g}, \\
    \frac{\partial h_\text{adh}}{\partial u_{l,-\bm b}^\perp} & = (-1)^l \frac{a}{L} \frac{\hat z \times \bm g}{ig^2} \cdot \left( \frac{\partial \mathcal V}{\partial \bm \phi} \right)_{\bm g},
\end{align}
\end{widetext}
with $\bm \phi = M \bm r + \bm u_1 - \bm u_2$ the local stacking. Using these equations, we then obtain the self-consistent equations shown in Eqs.\ \eqref{eq:sc1} and \eqref{eq:sc2} of the main text. In Fig.\ \ref{fig:app0}, we show the local stacking energy for the relaxed configuration for graphene on hBN. Numerical results for the Fourier components for graphene on hBN are shown in Fig.\ \ref{fig:app1} and for 2H TMD heterobilayers in in Fig.\ \ref{fig:app2}.

\subsection{Perturbation theory}

It is convenient to write the original layer-resolved fields in terms of the layer center-of-mass ($\bm U$) and relative displacements ($\bm u$). This is motivated by the fact that the latter is dominant, while the former is expected to be much smaller. We define these auxiliary fields as
\begin{equation}
    \bm u_{1,2} = \frac{\bm U \pm \bm u}{2}.
\end{equation}
We can rewrite the elastic energy in terms of these auxiliary fields. Because we only consider in-plane displacements, the strain tensor is linear,
\begin{equation}
    u_{ij}^{(1,2)} = \frac{U_{ij} \pm u_{ij}}{2},
\end{equation}
and
\begin{align}
    u_{ii}^{(1,2)} u_{ii}^{(1,2)} & = \frac{1}{4} \left( U_{ii} \pm u_{ii} \right) \left( U_{ii} \pm u_{ii} \right) \\
    & = \frac{1}{4} \left( U_{ii} U_{ii} + u_{ii} u_{ii} \pm 2 u_{ii} U_{ii} \right),
\end{align}
and similar for the second term in the elastic energy. We therefore obtain
\begin{equation}
    \begin{aligned}
        H_\text{elas} & = \frac{1}{4} \int d^2 \bm r \, \Big( \lambda U_{ii} U_{ii} + 2 \mu U_{ij} U_{ji} \\
        & + \lambda u_{ii} u_{ii} + 2 \mu u_{ij} u_{ji} + \bar \lambda u_{ii} U_{ii} + 2 \bar \mu u_{ij} U_{ji} \Big),
    \end{aligned}
\end{equation}
with effective Lam\'e parameters
\begin{alignat}{3}
    \lambda & = \frac{\lambda_1 + \lambda_2}{2}, \qquad && \mu && = \frac{\mu_1 + \mu_2}{2}, \\
    \bar \lambda & = \lambda_1 - \lambda_2, \qquad && \bar \mu && = \mu_1 - \mu_2.
\end{alignat}

To obtain approximate expressions for the relaxed configuration, we solve the self-consistent equations perturbatively in lowest order. Here, the small parameter is given by the ratio of the characteristic adhesion and elastic energy. We now make use of the fact that the acoustic displacement fields of the adiabatic relaxed ground-state configuration are constrained by the symmetry of the rigid moir\'e. These symmetry constraints (see Table \ref{tab:tab1}) give rise to the following first-star \textit{ansatze}:
\begin{widetext}
\begin{align}
    \bm u(\bm r) & = \frac{\sqrt{3}a}{2\pi} \sum_{n=1}^3 \left[ u_1^\parallel \hat g_n \sin( \bm g_n \cdot \bm r + \alpha_1^\parallel ) + u_1^\perp \hat z \times \hat g_n \sin( \bm g_n \cdot \bm r + \alpha_1^\perp ) \right], \\
    \bm U(\bm r) & = \frac{\sqrt{3}a}{2\pi} \sum_{n=1}^3 \left[ U_1^\parallel \hat g_n \sin( \bm g_n \cdot \bm r + \beta_1^\parallel ) + U_1^\perp \hat z \times \hat g_n \sin( \bm g_n \cdot \bm r + \beta_1^\perp ) \right],
\end{align}
where the subscript refers the first star of reciprocal vectors instead of the layer index. For example, we defined $u_{1,0}^\parallel = u_1^\parallel e^{i\alpha_1^\parallel}$, $u_{1,0}^\perp = u_1^\perp e^{i\alpha_1^\perp}$, and similarly for the center-of-mass displacement. If we plug these expression in the elastic and adhesion energy, we find
\begin{align}
    h_\text{elas} & = \frac{a^2}{L^2} \Big\{ \frac{3\mu}{2} \left[ ( U_1^\perp )^2 + ( u_1^\perp )^2 \right] + \frac{3 \bar \mu}{2} u_1^\perp U_1^\perp \cos( \alpha_1^\perp - \beta_1^\perp ) \\
    & + \frac{3\left( \lambda + 2 \mu \right)}{2} \left[ ( U_1^\parallel )^2 + ( u_1^\parallel )^2 \right] + \frac{3\left( \bar \lambda + 2 \bar \mu \right)}{2}  u_1^\parallel U_1^\parallel \cos( \alpha_1^\parallel - \beta_1^\parallel ) \Big\}, \\
    h_\text{adh} & = -\frac{6 \mathcal V_1 L}{a} \left[ u_1^\parallel \epsilon \cos( \alpha_1^\parallel - \varphi_1 ) + u_1^\perp \theta \cos( \alpha_1^\perp - \varphi_1 ) \right],
\end{align}
for small twist angles $M \approx \epsilon \sigma_0 - i \theta \sigma _y$. Minimizing the total energy yields $\alpha_1^\parallel = \alpha_1^\perp = \beta_1^\parallel = \beta_1^\perp = \varphi_1$ and
\begin{alignat}{3}
    u_1^\parallel & = \left( \frac{1}{\lambda_1 + 2 \mu_1} + \frac{1}{\lambda_2 + 2 \mu_2} \right) \frac{\mathcal V_1 \epsilon}{( \epsilon^2 + \theta^2 )^{3/2}}, \qquad && U_1^\parallel && = \left( \frac{1}{\lambda_1 + 2 \mu_1} - \frac{1}{\lambda_2 + 2 \mu_2} \right) \frac{\mathcal V_1 \epsilon}{( \epsilon^2 + \theta^2 )^{3/2}}, \label{eq:u1para} \\
    u_1^\perp & = \left( \frac{1}{\mu_1} + \frac{1}{\mu_2} \right) \frac{\mathcal V_1 \theta}{( \epsilon^2 + \theta^2 )^{3/2}}, \qquad && U_1^\perp && = \left( \frac{1}{\mu_1} - \frac{1}{\mu_2} \right) \frac{\mathcal V_1 \theta}{( \epsilon^2 + \theta^2 )^{3/2}}, \label{eq:u1perp}
\end{alignat}
\end{widetext}
which recovers the results from Refs.\ \cite{ezziPRL2024,ceferino2M2024} for homobilayers with $\epsilon \rightarrow 0$ and finite $\theta$.

\section{Buckling} \label{app:buckling}

\subsection{Elastic energy}

Including the out-of-plane bending term, the elastic energy of a single layer can be written as \cite{nelsondavid2004}
\begin{equation}
    H_\text{elas} = \frac{1}{2} \int d^2 \bm r \left[ \lambda u_{ii} u_{ii} + 2 \mu u_{ij} u_{ji} + \kappa \left( \nabla^2 h \right)^2 \right],
\end{equation}
with 2D Lam\'e parameters $\lambda$ and $\mu$, and where $\kappa$ is the out-of-plane bending rigidity. The strain tensor now also includes the quadratic out-of-plane part:
\begin{equation}
    u_{ij} = \frac{1}{2} \left( \frac{\partial u_j}{\partial r_j} + \frac{\partial u_i}{\partial r_j} + \frac{\partial h}{\partial r_i} \frac{\partial h}{\partial r_j} \right).
\end{equation}
Using the Helmholtz decomposition for a periodic in-plane displacement field (now without the prefactor $a/L$), we find that the elastic energy can be written in Fourier space as
\begin{widetext}
\begin{equation} \label{eq:elas2}
    \frac{1}{2} \sum_{\bm g} \left[ \left( \lambda + 2 \mu \right) \left| u_{\bm g}^\parallel + \frac{f_{xx\bm g} + f_{yy\bm g}}{2} \right|^2 + \mu \left( \frac{|u_{\bm g}^\perp|^2}{2} - \mathcal G_{-\bm g} u_{\bm g}^\perp - \mathcal F_{-\bm g} u_{\bm g}^\parallel + \frac{|f_{xy\bm g}|^2 - f_{xx\bm g} f_{yy,-\bm g}}{2} + \mathrm{c.c.} \right) + \kappa | h_{\bm g} |^2 g^4 \right],
\end{equation}
\end{widetext}
with $\mathcal F_{\bm g} = \left( g_x^2 f_{yy\bm g} - 2 g_x g_y f_{xy\bm g} + g_y^2 f_{xx\bm g} \right) / g^2$ which transforms as a scalar field, and is related to the Hessian determinant of the surface in Monge gauge. Here, we also defined the symmetric tensor $f_{ij}(\bm r) = (\partial_i h)(\partial_j h)$ which is the metric tensor up to the identity matrix, with
\begin{equation}\label{eq:f_ijg}
    f_{ij\bm g} = - \frac{1}{2} \sum_{\bm g'} h_{\bm g'} h_{\bm g - {\bm g}'} \left( g_i g_j' + g_j g_i' - 2 g_i' g_j' \right).
\end{equation}
Here, we also defined
\begin{equation} \label{eq:app1}
    \mathcal G_{\bm g} = \frac{\left( f_{xx\bm g} - f_{yy\bm g} \right) g_x g_y + f_{xy\bm g} \left( g_y^2 - g_x^2 \right)}{g^2}.
\end{equation}

As a check, we recover the results of Refs.\ \cite{guineaPRB2008,guineaSSC2009,wehlingE2008,phongPRL2022,debeulePRL2025} if we minimize with respect to $u_{-\bm g}^\parallel$ for fixed $h_{\bm g}$. This yields
\begin{align}
    u_{\bm g}^\parallel & = \frac{\mu \mathcal F_{\bm g}}{\lambda + 2\mu} - \frac{f_{xx\bm g} + f_{yy\bm g}}{2}, \\
    u_{\bm g}^\perp & = \mathcal G_{\bm g},
\end{align}
which agrees with Ref.\ \cite{debeulePRL2025}. 
Using the definition of $f_{ij\bm g}$, the right-hand side of Eq.\ \eqref{eq:app1} can be written as
\begin{equation}
    \mathcal G_{\bm g} = \frac{1}{g^2} \sum_{\bm g'} h_{\bm g'} h_{\bm g-\bm g'} \left( \bm g' \cdot \bm g \right) \left( \bm g' \times \bm g \right),
\end{equation}
where we define the cross product between two in-plane vectors as a pseudoscalar. Here, we used
\begin{equation}
    \sum_{\bm g'} h_{\bm g'} h_{\bm g-\bm g'} \left( \bm g' \times \bm g \right) = 0.
\end{equation}
Similarly, we find
\begin{equation}
    \mathcal F_{\bm g} = \frac{1}{g^2} \sum_{\bm g'} h_{\bm g'} h_{\bm g-\bm g'} \left( \bm g' \times \bm g \right)^2,
\end{equation}
which is related to the Hessian determinant:
\begin{equation}
    ( \partial_x^2 h ) ( \partial_y^2 h ) - ( \partial_x \partial_y h )^2 = \frac{1}{2} \sum_{\bm g} g^2 \mathcal F_{\bm g}.
\end{equation}

Unlike $\mathcal F_{\bm g}$ we see that $\mathcal G_{\bm g}$ transforms as a pseudoscalar under a symmetry $\mathcal S$ of the out-of-plane displacement field, i.e., for $h_{\bm g} = h_{\mathcal S \bm g}$. This is consistent with the transformation laws
\begin{equation}
    u_{\mathcal S \bm g}^\parallel = u_{\bm g}^\parallel, \qquad u_{\mathcal S \bm g}^\perp = \det(\mathcal S) u_{\bm g}^\perp.
\end{equation}
Hence, we find that $\mathcal F_{\bm g}$ and $\mathcal G_{\bm g}$ obey the same symmetry constraints as $u_{\bm g}^\parallel$ and $u_{\bm g}^\perp$, respectively. For a height profile with $\mathcal C_{3v}$ symmetry, we note that $\mathcal G_{\bm g}$ vanishes for the first-star because it transforms as a pseudoscalar:
\begin{equation}
    \mathcal G_{\mathcal M_x \bm g} = -\mathcal G_{\bm g},
\end{equation}
which implies that $\mathcal G_{\bm g} = 0$ for $\mathcal M_x \bm g = \bm g$. This is the case for any multiple of the first-star reciprocal vector $\bm g_1$. Hence, together with $\mathcal C_3$ this implies that for a first-star in-plane displacement field, and a buckling profile with $C_{3v}$ symmetry, there is no coupling to the rotational piece. However, this is only strictly true for the aligned (or anti-aligned) case, as the symmetry is reduced to $C_3$ when the twist angle is finite. We assume that these corrections, which are proportional to imaginary part of the second star ($\sin \gamma_2$) in lowest order, can be neglected.

\subsection{Buckling equation}

In order to find the out-of-plane displacement field under periodic boundary conditions for a given in-plane displacement field (specified by $u_{\bm g}^\parallel$ and $u_{\bm g}^\perp$), we minimize the elastic energy of a single layer with respect to $h_{-\bm g}$.
To this end, we first compute
\begin{align}
    \frac{\partial f_{ij,\bm g'}}{\partial h_{-\bm g}} & = \left[ g_i ( g_j + g_j' ) + g_j ( g_i + g_i' ) \right] h_{\bm g + \bm g'}, \\
    \frac{\partial \mathcal F_{\bm g'}}{\partial h_{-\bm g}} & = \frac{2 \left( \bm g \times \bm g' \right)^2}{g^{\prime 2}} \, h_{\bm g + \bm g'}, \\
    \frac{\partial \mathcal G_{\bm g'}}{\partial h_{-\bm g}} & = \frac{\left( \bm g \times \bm g' \right) \bm g' \cdot \left( 2 \bm g + \bm g' \right)}{g^{\prime 2}} \, h_{\bm g + \bm g'}.
\end{align}
Setting the partial derivative of the elastic energy of a single laver with respect to $h_{-\bm g}$ equal to zero then gives
\begin{widetext}
\begin{align}
    \kappa g^4 h_{\bm g} + \sum_{\bm g'} & \Bigg\{ -2\mu \Bigg[ \frac{2 \left( \bm g \times \bm g' \right)^2}{g^{\prime 2}} \, u_{\bm g'}^\parallel + \frac{\left( \bm g \times \bm g' \right) \bm g' \cdot \left( 2 \bm g - \bm g' \right)}{g^{\prime 2}} \, u_{\bm g'}^\perp \\
    & + g_x (g_x - g_x') f_{yy\bm g'} + g_y (g_y - g_y') f_{xx\bm g'} + \left( g_x g_y' + g_y g_x' - 2 g_x g_y \right) f_{xy\bm g'} \Bigg] \\
    & + \left( \lambda + 2 \mu \right) \bm g \cdot \left( \bm g - \bm g' \right) \left( u_{\bm g'}^\parallel + \frac{f_{xx\bm g'} + f_{yy\bm g'}}{2} \right) \Bigg\} h_{\bm g - \bm g'} = 0,
\end{align}
which has to be solved numerically in general.
\end{widetext}


\bibliography{references,refs-DFT}


\end{document}